\pdfoutput=1 
\documentclass{mn2e}  
\usepackage{graphicx}

\title{The shear-induced alpha-effect and long-term variations in
  solar dynamo}
\author[V.V.Pipin]{V.V.Pipin \thanks{E-mail:
pip@iszf.irk.ru}\\
Institute for Solar-Terrestrial Physics,
Siberian Division of Russian Academy of Sciences, 664033 Irkutsk,
Russia\\
}

\begin{document}

\date{Accepted \today. Received \today; in original form \today}

\pagerange{\pageref{firstpage}--\pageref{lastpage}} \pubyear{2008}

\maketitle

\begin{abstract}
The consequences of the shear-induced alpha effect to the long-term
modulation of magnetic activity are examined with the help of the
axisymmetric numerical dynamo model that includes the self-consistent
description of the angular momentum balance, heat transport and magnetic
field generation in the spherical shell. We find that the shear contributions
to alpha effect can complicate the long-term behaviour of the large-scale
magnetic activity and differential rotation in nonlinear dynamo. Additionally
we consider the impact secular magnetic activity variations to the
secular modulations of the solar luminosity and radius.
\end{abstract}

\section{Introduction}

It seems that the long-term variations of the sunspot's activity and
phenomena similar to the Maunder minimum are widely presented on the
late-type stars like the Sun, see e.g.\cite{soon94,bal90}. Generally,
it may be usefull to distinguish the different kinds of time variations
of magnetic activity on the stars for two type. One is irregular disappearence
of sunspot' cycle similarily to what was happened during the Maunder
minimum at the second half of the 17-th centure, \cite{eddy76}. Another
kind of variations can be attributed to the quasi-regular grand activity
cycles. The grand activity cycles is natural to explain as a result
of the nonlinear interactions between magnetic fields and the main
sources of the solar dynamo, which are the differential rotation and
convection, \cite{kit-ru-ku,weiss,tobias97,pip-glie,kuek99}. The
given nonlinear mechanisms may be responsible for the chaotic variations
of magnetic activity on the late type stars and on the Sun, as well.
Though, the chaotic behaviour was demostrated only in the case of
the highly supercritical dynamos, \cite{kit-ru-ku,tobias97,kuek99}.
Currently, some doubts are cast upon the possibilitiy for the supercritical
regimes in the stellar dynamo. The main constraint for feasibility
of the supercritical dynamo is due to magnetic helicity conservation.
However, the problem can be overcome if to allow the helicity outflux
of the dynamo region. Another contraint come into the play if to consider
the distributed dynamo. In this case the energy of the generated large-scale
magnetic fields (LSMF) is hardly to exceed the equipartition level
with the energy of convective flow. These arguments motivate us to
look for mechanisms which could induce the long-term variations for
such a weakly non-linear dynamo regimes.

In this paper we consider an effect that can provide a nonlinear coupling
between two main sources of the solar dynamo, these are the differential
rotation and $\alpha$ effect. In according to \cite{kuzetal05,pip05,pipk05}
the shear (differential rotation) changes both the structure and the
amplitude of the alpha effect. For the Sun the most important component
of alpha effect is $\alpha_{\phi\phi}$. The estimation of shear contribution
to the $\alpha_{\phi\phi}$ with regards fo the typical conditions
in the solar CZ shows that it gives only about 15\% of the whole magnitude.
In the nonlinear dynamo the LSMF are included in the angular momentum
balance inside CZ. In fact the grand activity cycle may be considered
as a result of relaxion of the pertubations angular momentum balance
in CZ emerged due to LSMF (via the $\Lambda$-quenching\cite{kit-ru-ku,pip-glie},
or the large-scale Lorenz force\cite{malcu-proct,tobias96,kuek99}).
In this case the incorporation of the shear-induced alpha effect can
complicate the relaxion process because it supplies the additional
destabilization the sources of magnetic filed generation due to the
time-dependence of the alpha effect on the differential rotation distribution
in CZ. 

Below, we demonstrate the significance of the shear-induced alpha-effect
for the long-term evolution of magnetic activity using the numerical
model of the axisymmetric dynamo in the convective shell. The model
includes the numerical solution for the dynamo equations, angular
momentum balance and heat transport problem with reagards for the
physical conditions that take place in the main part of the solar
CZ. In the next part of the paper we give the short description of
the model (more details can be found in the article\cite{pi04}).
The results of numerical simulations are discussed at the third part
of the paper.

\section{The model design}

The basic physical quantities under considerations are the induction
vector of the axisymmetric LSMF, $\mathbf{B}=\mathbf{e}_{\phi}B(r,\theta)+curl\left(\frac{\mathbf{e}_{\phi}A\left(r,\theta\right)}{r\sin\theta}\right)$,
the large-scale flow - differential rotation, $\mathbf{V}=\mathbf{e}_{\phi}r\sin\theta\,\Omega\left(r,\theta\right)$
(we neglect the meridional circulation, here), the mean specific entropy,
$S\left(r,\theta\right)$, and the pressure, $P\left(r,\theta\right)$.
The magnetic activity and differential rotation give rise to pertubations
of the thermodynamic stage of SCZ. The thermodynamic deviations from
reference model (taken on the base of solar interior model from Stix\cite{sti02})
is described with help of the mean entropy production equation. In
following to Pipin \& Kitchatinov\cite{pip-rad} we write it as
 \begin{eqnarray}
\rho T\frac{\partial S}{\partial t} & =- & \frac{1}{r^{2}}\frac{\partial}{\partial r}r^{2}F_{r}-\frac{1}{r\sin\theta}\frac{\partial}{\partial\theta}\sin^{2}\theta F_{\theta}\label{seq}\\
 & + & \left(T_{\theta\phi}\frac{\partial\Omega}{\partial\theta}+rT_{r\phi}\frac{\partial\Omega}{\partial r}\right)\sin\theta-\frac{1}{4\pi}\left(\mathbf{\mathcal{E}\cdot rotB}\right),\nonumber \end{eqnarray}
Again, for details please look at paper\cite{pi04} and references
there. The heat flux $\mathbf{F}$ is a sum of convective and radiative
fluxes $\mathbf{F}=\mathbf{F_{c}}+\mathbf{F}_{rad},$ where in convective
flux we take into account both the influence of rotation and magnetic
fileds. The second string in (\ref{seq}) is responsible for the heating
and cooling of the matter in CZ due to generation and dissipation
of large-scale magnetic fileds and differential rotation there. 

The angular momentum transport in convective shell is described with,
\begin{eqnarray}
\rho r\sin\theta\frac{\partial\Omega}{\partial t} & = & \frac{1}{r^{3}}\frac{\partial}{\partial r}r^{3}\left(T_{r\phi}+\frac{F_{L}B}{4\pi r^{2}\sin\theta}\frac{\partial A}{\partial\theta}\right)\label{eq:ang-mom}\\
 & + & \frac{1}{r\sin^{2}\theta}\frac{\partial}{\partial\theta}\sin^{2}\theta\left(T_{\theta\phi}-\frac{F_{L}B}{4\pi r\sin\theta}\frac{\partial A}{\partial r}\right),\nonumber \end{eqnarray}

\begin{eqnarray*}
T_{r\phi} & = & \rho P_{\nu}\chi_{T}\sin\theta\biggl[\Phi_{\perp}\Psi_{1}\left(\Omega^{*},\beta\right)r\frac{\partial\Omega}{\partial r}\\
 & + &
 \left(\Phi_{\Vert}\Psi_{2}\left(\Omega^{*},\beta\right)
-\Phi_{\perp}\Psi_{2}\left(\Omega^{*},\beta\right)\right)\times 
 \\ 
&\times&
 \cos\theta\left(r\cos\theta\frac{\partial\Omega}
{\partial r}-\sin\theta\frac{\partial\Omega}{\partial\theta}\right)\\
 & - & \left(\frac{\alpha_{M}}{\gamma}\right)^{2}\Omega\left(\mathcal{J}_{0}-\mathcal{J}_{1}+c_{V}\sin^{2}\theta\mathcal{J}_{1}\right)\phi_{V}\left(\Omega^{*},\beta\right)\biggr]\\
T_{\theta\phi} & = & \rho P_{\nu}\chi_{T}\sin\theta\biggl[\Phi_{\perp}\Psi_{1}\left(\Omega^{*},\beta\right)\sin\theta\frac{\partial\Omega}{\partial\theta}\\
 & + & \left(\Phi_{\Vert}\Psi_{2}\left(\Omega^{*},\beta\right)
-\Phi_{\perp}\Psi_{1}\left(\Omega^{*},\beta\right)\right) \times \\
&\times& \sin^{2}\theta\left(\sin\theta\frac{\partial\Omega}{\partial\theta}-r\cos\theta\frac{\partial\Omega}{\partial r}\right)\\
 &-\!\! &\!\! \left(\frac{\alpha_{M}}{\gamma}\right)^{2}\!\!\Omega\cos\theta\!\!
\left(\!\!\mathcal{J}_{2}\phi_{H}\!\!\left(\Omega^{*},\beta\right)\!\!
+\!\!c_{V}\sin^{2}\theta\mathcal{J}_{1}\phi_{V}\!\!\left(\Omega^{*},\beta\right)\right)
\!\!\biggr].
\end{eqnarray*}
The turbulent stresses, $T_{(r,\theta)\phi}$, describe the effective
redistribution of the angular momentum in rotating convective shell
due to convection. The mesure of rotational influence on convection
is the Coriolis number, $\Omega^{*}=2\Omega_{0}\tau_{c}$; here $\Omega_{0}$is
the basic rotation rate of a star, and $\tau_{c}$is the typical convective
turnover time. The parameter $\beta$ mesures the influence of magnetic
filed on convection. It is defined as $\beta=\left|B\right|/\left(\sqrt{4\pi\rho}u_{c}\right)$,
where $u_{c}$ is rms convective velocity. Both $\tau_{c}$ and $u_{c}$
are taken from the reference solar interior model by Stix(2002). Functions,
$\Phi_{(\perp,\|)}$, $\mathcal{J}_{\left(1,2\right)}$ are responsible
for the influence of rotation on the turbulent viscosity and on the
$\Lambda$ effect repectively. They are given in papers\cite{kkit-rud-dr(93),kit-visc}.
The effect of magnetic fileds on viscosity and on the $\Lambda$ effect
(which is for breivity call as $\Lambda$-quenching) is approximated
with help of $\Psi_{1,2}$ and with $\psi_{(V,H)}$ respectively.
They are defined as follows. Its definitions can be found in appendix.
In case of the slow rotation we have $\Psi_{(1,2)}\left(\Omega^{*},\beta\right)=\psi_{\left(1,2\right)}\left(\beta\right)$,
$\psi_{(V,H)}\left(\Omega^{*},\beta\right)=K\left(\beta\right)_{(1,2)}$where
$\psi_{\left(1,2\right)}$ are given in ... and $K_{(1,2)}$- in ...
The $c_{V}$ is a paramer to adjust the differential rotation law
in shell in more close argreement with helioseismology results. We
use $c_{V}=1.3$ to get nearly radial profiles of angular velosity
isolines in the main part of CZ (see below). $P_{\nu}={\displaystyle \frac{\nu_{T}}{\chi_{T}}}$
is the effective Prandtl number and $\chi_{T}$, $\nu_{T}$ - typical
eddy heat conductivity and eddy viscosity respectively. In simulations
presetnted here we used $P_{\nu}=0.3,0.1,0.05$. These gives examples
for two different regimes. The effects of $\Lambda$-quenching prevails
the Lorentz force for $P_{\nu}=0.3$ and the opposite case is for
$P_{\nu}=0.05$. Aslo we have to introduce the parameter $F_{L}<1$
to tune the power of the large-scale Lorentz force. Its necessity
is determined by shortcomings in dynamo model. Particularily, in our
model the relation between the typical strength of toroidal and poloidal
components of the large-scale magnetic field is $B_{T}/B_{P}\approx50$
which is less than can be expected from observations, where $B_{T}/B_{P}>100$. 

We use the following dynamo equations in a spherical coordinate system, 

\begin{eqnarray}
\frac{\partial B}{\partial t} & = & \frac{1}{r}\left\{ \frac{\partial\left(\Omega,A\right)}{\partial\left(r,\theta\right)}+\frac{\partial r\mathcal{E}_{\theta}}{\partial r}-\frac{\partial\mathcal{E}_{r}}{\partial\theta}\right\} ,\label{beq_03}\\
\mathcal{E}_{r} & = & \chi_{T}P'_{m}\Biggl[B\left(-\frac{2\alpha_{M}\phi_{1}}{\ell_{c}\gamma}+\frac{\psi_{\eta}\phi_{\|}}{r}\right)\sin\theta\cos\theta\nonumber \\
 & - & \frac{\psi_{\eta}\left(\phi+\phi_{\|}\sin^{2}\theta\right)}{{r\sin\theta}}\frac{\partial\sin\theta B}{\partial\theta}+\phi_{1}\psi_{\eta}\sin\theta\cos\theta\frac{\partial B}{\partial r}\nonumber \\
 & + &
 C_{\alpha}\frac{\alpha_{M}\phi_{\alpha}}{\ell_{c}\gamma}\bigl(\left(\mathcal{A}_{1}
+\mathcal{A}_{2}+\mathcal{A}_{4}\cos^{2}\theta\right)\frac{\cos\theta}{r^{2}\sin\theta}
\frac{\partial A}{\partial\theta}\nonumber\\
&+&\left(\mathcal{A}_{2}+\mathcal{A}_{4}\cos^{2}\theta\right)\frac{1}{r}\frac{\partial
  A}{\partial r}
\bigr)\Biggr],\nonumber \end{eqnarray}

\begin{eqnarray*}
r\mathcal{E}_{\theta} & = & \chi_{T}P'_{m}\Biggl[B\left(r\left(\mathrm{\phi_{2}}-2\phi_{2}\sin^{2}\theta\right)\frac{\alpha_{M}}{\ell_{c}\gamma}-\phi_{\|}\psi_{\eta}\cos^{2}\theta\right)\\
 & + & \left.\psi_{\eta}\left(\phi+\phi_{\|}\cos^{2}\theta\right)\frac{\partial rB}{\partial r}-\psi_{\eta}\phi_{\|}\sin\theta\cos\theta\frac{\partial B}{\partial\theta}\right)\\
 & + &
 C_{\alpha}\frac{\alpha_{M}\phi_{\alpha}}{\ell_{c}\gamma}\bigl(\left(\mathcal{A}_{2}
+\mathcal{A}_{4}\cos^{2}\theta\right)\frac{1}{r^{2}}\frac{\partial
A}{\partial\theta}\\
&+&\left(\mathcal{A}{}_{1}+\mathcal{A}_{4}\sin^{2}\theta\right)\frac{\cos\theta}{r\sin\theta}
\frac{\partial A}{\partial r}\bigr)\Biggr],\end{eqnarray*}

\begin{eqnarray}
\frac{\partial A}{\partial t} & = &
\chi_{T}P'_{m}\psi_{\eta}\Biggl[\left(\phi
+\phi_{\|}\cos^{2}\theta\right)\frac{\partial^{2}A}{\partial r^{2}}\label{aeq_03}\\
&+&\frac{\phi+\phi_{\|}\sin^{2}\theta}{r^{2}}
\frac{\partial}{\partial\theta}\frac{1}{\sin\theta}\frac{\partial
  A}{\partial\theta}+\nonumber\\
 & + & \phi_{\|}\left(\frac{\sin^{2}\theta}{r}\frac{\partial A}{\partial r}+\frac{3\sin\theta\cos\theta}{r^{2}}\frac{\partial A}{\partial\theta}-\frac{\sin2\theta}{r}\frac{\partial^{2}A}{\partial r\partial\theta}\right)\nonumber \\
 & - &
 \frac{\alpha_{M}}{\ell_{c}\gamma}\bigl(\left(\cos^{2}\theta\left(\phi_{1}
-\phi_{3}\right)+\phi_{1}+\phi_{2}\right)\frac{\partial A}{\partial
r}\nonumber \\
&-&(\phi_{1}-\phi_{3})\frac{\sin2\theta}{2r}\frac{\partial A}{\partial\theta}\bigr)\Biggr] + 
 C{}_{\alpha}\chi_{T}P'_{m}\frac{\alpha_{M}}{\ell_{c}\gamma}\psi_{\alpha}\Biggl[\mathcal{A}_{1}\cos\theta \nonumber \\
&+ &\Psi_{\alpha1}\cos\theta\sin^{2}\left(\theta\right)\frac{\partial\log\left(\Omega\right)}{\partial\log\left(r\right)}\nonumber \\
 & + & \sin\left(\theta\right)\frac{\partial\log\left(\Omega\right)}{\partial\theta}\left(\Psi_{\alpha2}+\cos^{2}\left(\theta\right)\Psi_{\alpha1}\right)\Biggl]r\sin\theta B,\nonumber \end{eqnarray}
More details about the given dynamo equations can be found in ...
Below, we make some commentaries about (\ref{beq_03},\ref{aeq_03}).
The power of generation sources due to alpha effect is characterized
via $C_{\alpha}={\displaystyle \frac{\eta_{0}}{\alpha_{0}R_{\odot}}}$,
where $\eta_{0}$and $\alpha_{0}$ are the maxima of radial distributions
of magnetic diffusivity and alpha effect in CZ; $\alpha_{M}={\displaystyle \frac{\ell_{c}}{H_{p}}}$
is a mixing-length paramer, which relates the pressure scale hieght
$H_{p}$and typical mixing length of convective flows $\ell_{c}$;
$P'_{m}$ relates the eddy diffusivity and eddy heat conductivity
$\eta_{T}=P'_{m}\chi_{T}$. To match the period cycle we used $P'_{m}=0.05$
in what follows. Functions $\phi_{n,\|},\,\mathcal{A}_{n}$and $\Psi_{\alpha n}$
surve to determine the dependence of the turbulent drift, dissipation
and generation of magnetic fields on the Coriolis number. They are
discussed in ...{[}{]}. The magnetic quenching function for magnetic
diffusivity and transport effects is $\psi_{\eta}$. The $\alpha$
quenching is described with $\phi_{\alpha}$. Both are given in appendix.
The eddy viscosity is determined with help of mixing-length relations
for rms convective velocity $u_{c}$, $\chi_{T}={\displaystyle \frac{u_{c}\ell_{c}}{3}}$,
$u_{c}={\displaystyle \frac{\ell_{c}}{2}\sqrt{-\frac{g}{c_{p}}\frac{\partial S}{\partial r}}}$.

\subsection{Boundary conditions}

The system (\ref{eq:ang-mom},\ref{beq_03},\ref{aeq_03},\ref{seq})
is integrate on the volume domain which is confined on the radius
in $r=\left[0.715-0.96\right]R_{\odot}$ and on the polar angle -
from pole to pole. For the angular momentum balace problem we define
the stress-free conditions both at the bottom and at the top of CZ
via $T_{r\phi}+{\displaystyle \frac{B}{4\pi r^{2}\sin\theta}\frac{\partial A}{\partial\theta}}=0$.
For dynamo equations, at the bottom of CZ we put a superconductivity
condition. At the surface we assume the condition of the {}``partial
reflection'' of magnetic energy suggested previously by Kitchatinov
et al. .., i.e., for vector potential $A$ the vaccum conditions is
applied and the asimuthal filed B is related with poloidal component
of electromotive force via $\mathcal{E}_{\theta}=-{\displaystyle \frac{\eta_{T}B^{2}}{R_{\odot}B_{(0)}}}$.,
where we take$B_{(0)}=200$Gs. For the heat transfer problem, at the
bottom we take the vanishing convective flux and at the surface the
thermal flux is approximated an effective blackbody radiation: $F_{r}={\displaystyle \frac{\mathcal{L}_{\odot}}{4\pi r^{2}}\left(1+4\frac{T_{e}\delta S}{c_{p}T_{eff}}\right)}$,
where $T_{eff}$ is the the effectictive temperature of the photosphere
and $T_{e}$ is the temperature at the outer boundary of the integration
domain. The reference stratification inside domain is assumed nearly
adiabatic. It is fixed if to specify the temperature $T_{e}$and density
$\rho_{e}$at the outer boundary. We use the parameters from the solar
interior model by Stix, $T_{e}=1.93\times10^{5}K$and $\rho_{e}=3.6\times10^{-3}$g\textasciitilde{}cm$^{-3}$.

\section{Results and discussion.}

Before discussion results it remains to fix some of parameteres introduced
above. The mixing-length parameter is specified to $\alpha_{M}=1.3$.
The low magnetic Prandtl number $P_{m}=0.05$ helps to match the period
of magnetic cycle in the model. Note, the maximum of radial heat conductivity
in our model is $\chi_{T}=2.1\times10^{13}$cm$^{3}$s$^{-1}$so the
correspondent magnetic diffusivity is about $10^{12}$cm$^{3}$s$^{-1}$.
The threeshold $C_{\alpha}$was obtained via numerical trials. As
an intial condition we choose the weak magnetic field with zero parity
index (see the definition below). The dipole parity activates first
for $C_{\alpha}^{cr}\geq0.66$. For the Prandtl number $P_{\nu}=1$
we did not find the long cycle within range $0.66<C_{\alpha}<1$.
Note, the choice $C_{\alpha}\geq1$ would mean that we are going to
inject more energy in alpha-effect than it is available from convective
flows. The long cycles activate after increase the hydrodynamical
response time on the angular momentum balance pertubations by magnetic
fields. For analyzing purpose it is useful to introduce the following
global quantities of solution. The overall magnetic activity is quantified
by integral of modulus of magnetic flux contained in CZ, $F_{M}=\int_{0}^{\pi}\int_{r_{b}}^{r_{e}}\left|B\left(r,\theta\right)\right|r\sin\theta\mathbf{d}r\mathbf{d}\theta$
The parity index - $P={\displaystyle \frac{E^{(S)}-E^{(A)}}{E^{(S)}+E^{(A)}}}$,
where $E^{(S)}$, $E^{(A)}$ the integrated over convection zone energies
of the quadrupole and dipole components of magnetic fields, respectively. 

The main subject of the paper concerns with the influence of the shear
contributions to alpha effect on the long-term variations in solar
dynamo. The system (\ref{eq:ang-mom},\ref{beq_03},\ref{aeq_03},\ref{seq})
forms the solid self-consitent basis for investigation this problem
with numerical simulations. Previous studies show that interaction
of the LSMF and differential rotation is, perhaps, the most important
factor that is responsible for excitation of the secular magnetic
activity cycles in the non-linear stellar dynamo. Coupling the LSMF
generation and angular momentum transport is described by (\ref{eq:ang-mom},\ref{beq_03},\ref{aeq_03}).
The changes of the heat transport in CZ due to LSMF activity and differential
rotation result to modifications both in the dynamo processes and
in the angular momentum balance. The influnce of these modifications
to the secular magnetic activity variations needs the separate study.
We will not address this problem here. At the end we show some preliminary
results about influence of the LSMF and differential rotation variations
on the changes in solar irradiance and radius. Just below we consider
results where the heat transport problem (\ref{seq}) is fixed at
the beginning of time-stepping procedure by solving (\ref{eq:ang-mom},\ref{seq})
without regards for the LSMF. 

\begin{figure*}
\begin{center}\includegraphics[%
  width=9cm,
  height=13cm]{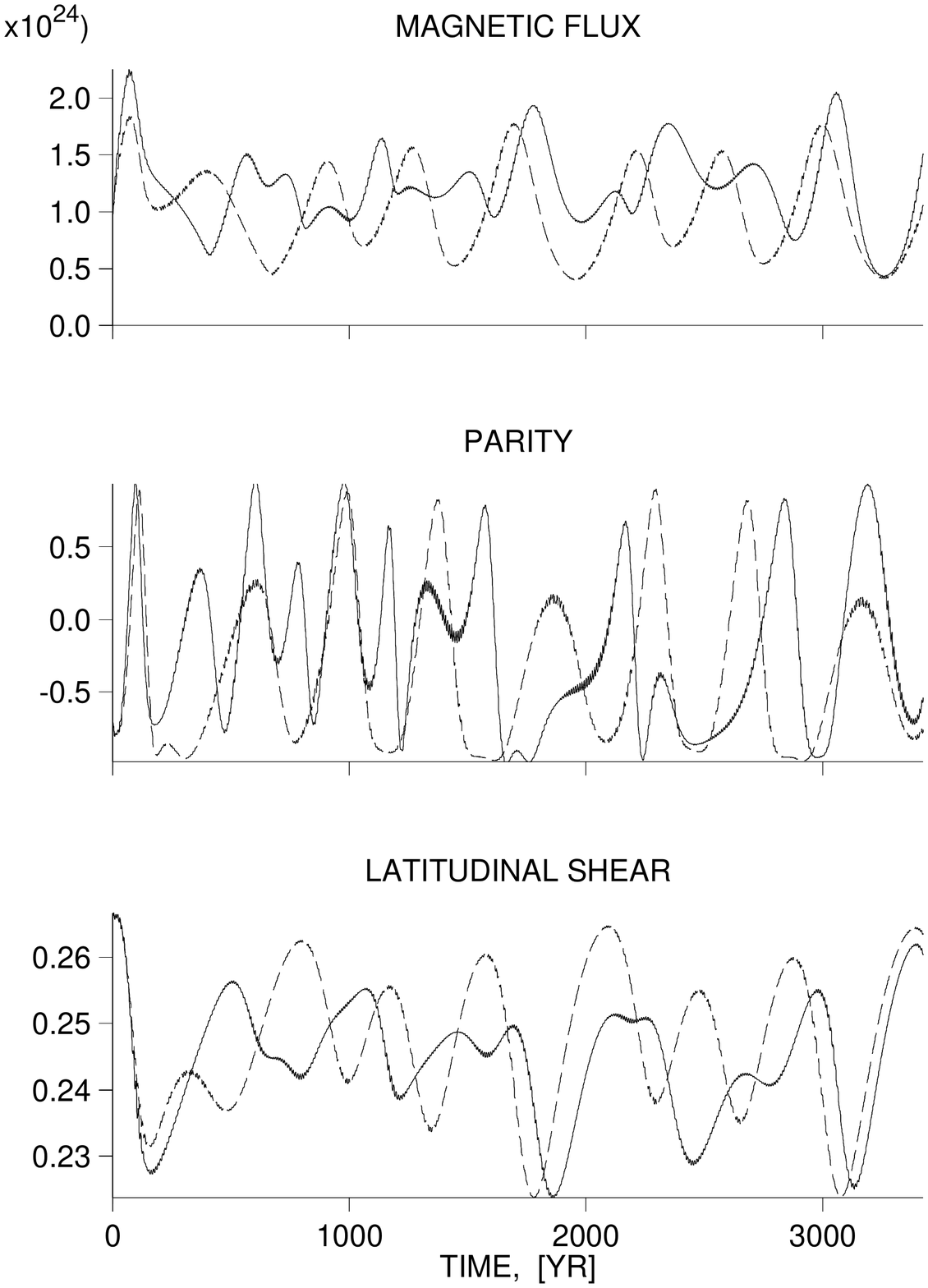}\includegraphics[%
  width=9cm,
  height=13cm]{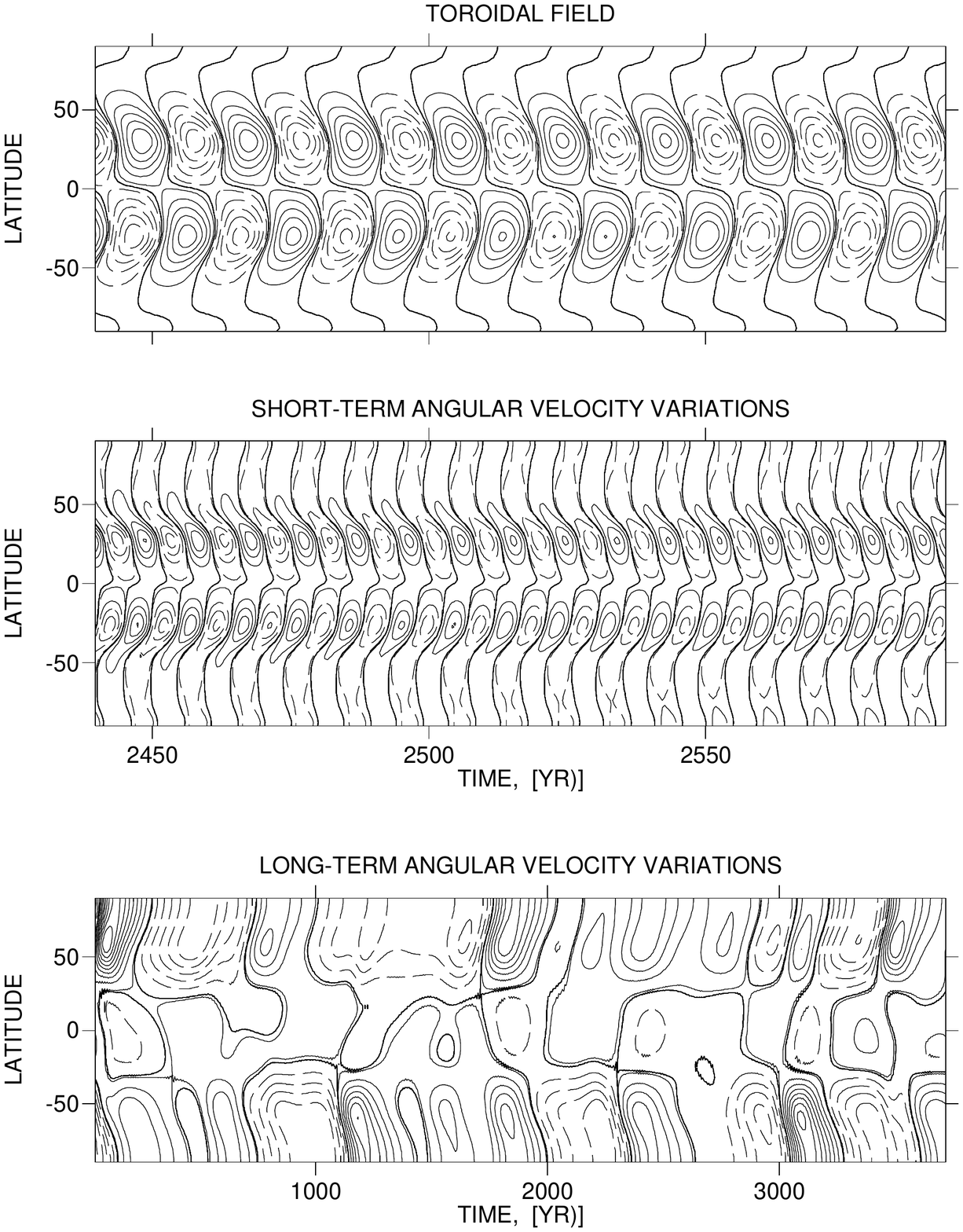}\end{center}

\caption{\label{cap:lt1}Variations of magnetic flux and parity (lelt) and
the Maunder diagramms for toroidal magnetic fields and angular velocity
variations (right).}
\end{figure*}
The Fig.\ref{cap:lt1} shows some results of numerical simulations
in the case of $C_{\alpha}=0.85$. On the left side of the Fig.\ref{cap:lt1}
we put the results concerning the variations of the integral magnetic
flux $F_{M}$(top), parity $P$(middle) and latitudinal shear (bottom).
The shear is quantified by the difference of angular velocity between
equator and poles. This difference is averaged over N and S-hemispheres.
There with solid line we draw the result of simulated evolutions with
regards for shear-induced alpha effect. The dashed line shows the
same but with the discarded contribution of shear to alpha effect.
In the picture we filtered out the short cycle to see the long-term
behaviour more clearly. In both case we see the long-term variation
of the magnetic activity. They are caused (similar to \cite{pip-glie})
primerely by the global interaction of LSMF and differential rotation.
It is clearly seen that pattern of the long-term variations is much
more complicated if the shear-induced alpha effect is allowed for.
Though this effect is certainly not the principal factor in exciting
the long-term variation. In particular, for this model the most important
factors driving the long-term variations are the $\Lambda$-quenching,
the large-scale Lorentz force and the non-linear magnetic diffusion.
The long-term variations disappears if one of them is excluded.

On the right side of the Fig.\ref{cap:lt1} we draw the butterfly
diagramm of toroidal LSMF near the surface for time interval about
200 years (top). The amplitude of magnetic field changes is 400 Gs.
The corresponding diagramm of the torsional variations of angular
velocity at the surface for this period is given at the middle. The
variations is measured by deviations of the current value of angular
velocity from the moving mean quantity obtained with averaging on
the period of cycle. The frequency of torsional variations is twice
to the basic frequency of magnetic cycle. The magnitude of the deviations
of roatational velocity is about 4$m/s$. The diagramm of the long-term
variations of angular velocity spanning the whole interval of simulations
is at the bottom. The short-term torsional oscillations are filtered
out there. The negative deviations of velocity are shown with dashed
contours and positive ones are shown with solid lines. The amplitude
of seqular modulation of surface angular velocity is about 40 m/s.
According to the given diagramm the relatively accelerated (or slow
down) zones occupy the high latitude of the solar surface and they
drift to poles in course of grand activity cycle. On the picture some
isolines span from pole to pole what can be interpreted as the angular
momentum exchanges between hemispheres. This confirms the global nature
of the long-term variations demonstrated by our model.

\begin{figure*}
\begin{center}\includegraphics[%
  width=13cm,
  height=10cm]{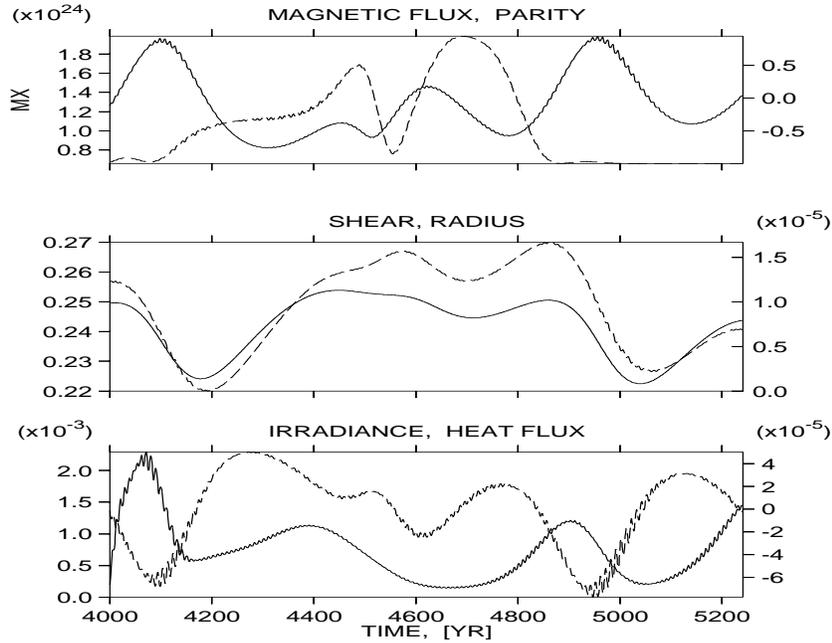}\end{center}

\caption{\label{cap:lt2}Long-term variations of the integral magnetic flux
(top, solid line, left scale), parity (top, dashed line, right scale),
latitudinal shear (middle, solid line, left scale), relative radius
variations (middle, dashed line, right scale), relative variations
of luminocity (bottom, solid line, left scale) and the relative heat
outflux variations (bottom, dashed line, right scale).}
\end{figure*}

At the end we discuss briefly the possible impact of the long-term
variations of magnetic activity and differential rotation to the heat
transport and hydrostatic balance in SCZ. The heat transport is described
with eq.(\ref{seq}). The description of hydrostatic balance with
regards for LSMF and differential rotation is given in , .The contribution
of magnetic field to irradiance output is estimated by the outflux
of the LSMF's energy from the dynamo region. This gives the maximum
amplitude of variations in solar irradiance if we assume that the
energy of escaping field is completely converted to radiation. The
radius variations were estimated from the changes of the zero mode
of the gravitational potential. On the Fig.\ref{cap:lt2} we show
of the long-term variations of the integral parameters describing
the integral magnetic flux and parity (top), the latidudinal shear
and the relative radius variations(middle), the relative variations
of luminocity and the heat outflux (bottom). The short cycles were
fitered out there. The panel in the middle of Fig.\ref{cap:lt2} clearly
shows that variations of the latitudinal shear and solar radius go
in phase, while they vary slightly ahead the changes of the magnetic
activity. In our model the delay is about 100 years. It worth to note
that in the short, 11-th year cycles the radius varies nearly in phase
with magnetic cycle with relative amplitude about $10^{-6}R_{\odot}$.
The amplitude of secular variations of radius is larger but it still
rather small, $1.5\times10^{-5}R_{\odot}$. This model seems to confirm
our previus result. In .. we concluded that the primary source of
the solar radius changes could be the modulation of the centrifugal
forces by magnetic activity (via modulation of differential rotation).
In according to simulations the secular modulation of luminocity due
to the escape of the magnetic energy from the dynamo region is about
$2\times10^{-3}\mathcal{L}_{\odot}$. This is the \emph{maximum possible}
estimate for the LSMF input to radiation. The seqular variations of
the heat outflux in the model are rather small,$10^{-4}\mathcal{L}_{\odot}$.
They are in anti-phase relation with magnetic activity. 

In conclusion we would like to summaries the results of the paper
as follows. The differential rotation modifies the standard alpha
effect changing its amplitude and inducing the additional mean electromotive
force that is perpendecular to original magnetic field. The dependence
of the alpha effect on differential can complicate the long-term variations
in non-linear dynamo. The main shortcomming of the model presented
is that we ignored the helicity conservation law. Some preliminary
results given in .. show that the helicity conservation could controll
the largest time scale of long-term variation in the non-linear dynamo.
The present model demonstrates the secular variations of radius. The
secular maximum of the radius corresponds to epoch the maximal differential
rotation and it foregoes the secular maximum of magnetic activity.
The amplitude of of secular variations of radius is $1.5\times10^{-5}R_{\odot}$.
The corresponded variations of solar irradiance are about $2\times10^{-3}\mathcal{L}_{\odot}$.
Given results suggest the futher work in directionWe hope that the
futher development of the dynamo theory and the model elaboration
will allow to make more solid conclusions about influence of the nonlinear
dynamo pr changes of the global parameters of the Sun refinement

\end{document}